\documentclass[reprint, amsmath, amssymb, aps, prl, superscriptaddress]{revtex4-2}

\usepackage{graphicx}
\usepackage{dcolumn}
\usepackage{bm}
\usepackage{hyperref}
\usepackage{color}

\begin{document}

\title{Universal Geometric Scaling in Cosmic Ray Spallation: \\ Evidence of a Dynamical Causal Horizon from AMS-02}

\author{Yi Yang}
\email{yiyang429@as.edu.tw}
\affiliation{Institute of Physics, Academia Sinica, Taipei 11529, Taiwan}
\affiliation{Department of Physics, National Cheng Kung University, Tainan 70101, Taiwan}

\date{\today}

\begin{abstract}
The interpretation of high-precision cosmic ray spectra is fundamentally bottlenecked by uncertainties in fragmentation cross-sections. Traditional kinematic models, driven by phase-space expansions, typically predict complex, energy-dependent evolutions. However, AMS-02 measurements reveal that at high rigidities ($R > 30$~GV), secondary-to-secondary flux ratios (Li/B, Be/B, and Li/Be) strictly converge to energy-independent plateaus. To understand this anomaly, we explore a macroscopic geometric framework. The ultra-relativistic spallation of a target nucleus snaps residual strong-interaction flux tubes, inducing an extreme deceleration on the remnant. Using a semi-microscopic estimation based on the Woods-Saxon potential and pion exchange, we suggest this dynamically generates a causal horizon with an effective Unruh temperature $T_U \approx 5.6-5.8$~MeV. Utilizing the Be/B ratio as an absolute calibration channel, we extract an asymptotic scale of $6.08$~MeV, remarkably consistent with our theoretical estimation and the established nuclear liquid-gas phase transition limit. Subsequent blind tests on Lithium ratios demonstrate a universal zero-slope convergence, providing compelling evidence that a constant geometric thermal bath effectively supersedes complex microscopic kinematics at high energies.
\end{abstract}

\maketitle

\textit{Introduction.}---The Alpha Magnetic Spectrometer (AMS-02) has propelled cosmic ray (CR) astrophysics into a high-precision era. While primary and secondary nuclei fluxes are measured to $\sim 1\%$ accuracy up to multi-TV rigidities~\cite{AMS_LiBeB, AMS_BC}, the theoretical interpretation of interstellar propagation remains severely constrained by $\mathcal{O}(10-20\%)$ uncertainties in microscopic fragmentation cross-sections ($\sigma_{\text{frag}}$). Standard propagation frameworks, most notably the GALPROP code~\cite{GALPROP}, rely heavily on conventional empirical parameterizations (e.g., Webber or Silberberg-Tsao~\cite{Webber2003}) that extrapolate low-energy accelerator data. 

In standard kinematic frameworks, production cross-sections are governed by multi-particle phase-space integrals ($\int d^3p / E$). At higher energies, the progressive opening of multi-pion channels forces these phase spaces to inherently grow logarithmically ($\ln s$). To avoid predicting gradual, channel-dependent logarithmic drifts in cross-section ratios, traditional empirical models must ad-hoc invoke the "hypothesis of limiting fragmentation"~\cite{LimitingFrag} to mathematically force asymptotic flatness at high energies. However, a fundamental, first-principles mechanism dictating exactly \textit{why} these cross-sections decouple from expanding phase spaces, and \textit{what} physical scale governs their asymptotic values, has remained elusive.

This underlying mechanism is uniquely illuminated by recent AMS-02 observations. Since secondary nuclei (Li, Be, B) share identical interstellar propagation histories, taking the ratio of two secondary fluxes analytically cancels the galactic escape grammage $\tau_{\text{esc}}(R) \propto R^{-\delta}$. For rigidities $R > 30$~GV, where solar modulation is negligible, this ratio uniquely isolates the pure microscopic production cross-sections. Contrary to pure phase-space expectations, AMS-02 secondary ratios exhibit remarkably flat, zero-slope plateaus~\cite{AMS_LiBeB}. 

In this Letter, we demonstrate that this high-energy "limiting fragmentation" behavior is a natural manifestation of macroscopic spacetime geometry. Building upon the geometric thermalization concepts discussed for heavy-ion Quark-Gluon Plasmas (QGP)~\cite{Castorina2007, Yang_PaperI}, we construct a causal decoupling framework for high-energy $p+A$ spallation, suggesting that asymptotic fragmentation yields are well-described by an effective Unruh thermal bath defined by intrinsic nuclear geometry.

\textit{Semi-Microscopic Estimation of the Nuclear Horizon.}---In low-energy spallation, fragmentation is governed by statistical thermal equilibration over a prolonged intra-nuclear cascade. However, at extreme rigidities ($R > 30$~GV), the incident proton traverses the target nucleus in a Lorentz-contracted timescale $\tau_{\text{coll}} \sim 2R_A/\gamma c \ll 0.1$~fm/$c$, which is vastly shorter than the typical nuclear relaxation time. During this ultra-relativistic penetration, the target nucleus is violently sheared, stretching the residual strong-interaction flux tube (mediated primarily by pion exchange) between the stripped fast nucleons and the spectator remnant.

The rapid snapping of this nuclear string induces an extreme proper deceleration ($a$) on the remnant. We provide a physically motivated semi-microscopic estimate for this deceleration. The restoring force exerted on a nucleon being pulled out of the nucleus is the gradient of the mean-field Woods-Saxon potential: $F = |dV/dr|$. This force reaches its absolute maximum at the nuclear surface, defining an effective string tension:
\begin{equation}
    F_{\text{max}} = \left| \frac{d}{dr} \frac{-V_0}{1 + \exp(\frac{r-R}{a_0})} \right|_{\text{max}} = \frac{V_0}{4a_0}.
\end{equation}
Using standard nuclear parameters (potential depth $V_0 \approx 50$~MeV, surface skin thickness $a_0 \approx 0.5$~fm), the maximum tension is $F_{\text{max}} \approx 25$~MeV/fm. Converting to natural units ($\hbar c \approx 197.3$~MeV$\cdot$fm), this corresponds to $F_{\text{max}} \approx 4932$~MeV$^2$.

Assuming the snapping of this residual force string is characterized by the production of the relevant exchange quantum, the pion (incorporating both neutral and charged pion masses, $m_\pi \approx 135-140$~MeV), the equivalent proper acceleration of the remnant string endpoint is $a = F_{\text{max}} / m_\pi$. According to the equivalence principle, this extreme deceleration dynamically generates a causal event horizon in the comoving frame~\cite{Castorina2007}, radiating with an Unruh temperature $T_U = a / 2\pi$. We estimate:
\begin{equation}
    T_U = \frac{V_0}{8\pi a_0 m_\pi} \approx \frac{4932 \text{ MeV}^2}{2\pi (140 \text{ MeV})} \approx 5.6 \sim 5.8 \text{ MeV}.
    \label{eq:prediction}
\end{equation}

Once this horizon forms, it induces a causal decoupling. Rather than explicitly tracking complex microscopic decay networks and sequentially expanding phase spaces, the production of fragments can be approximated as a quantum tunneling process across the horizon. We propose a minimal geometric ansatz for the asymptotic cross-sections, weighted by the separation energy penalty $\Delta Q$:
\begin{equation}
    \frac{\sigma_{i}}{\sigma_{j}} \approx \left( \frac{\Omega_i}{\Omega_j} \right) \exp\left( -\frac{Q_{i} - Q_{j}}{T_U} \right),
    \label{eq:ansatz}
\end{equation}
where $\Omega$ acts as a phenomenological combinatorial factor. Crucially, because $T_U$ is determined by intrinsic, energy-independent nuclear properties ($V_0, a_0, m_\pi$) rather than collision kinematics, it enforces flat high-energy plateaus.

\textit{Calibration via the Golden Channel.}---We test this geometric interpretation against AMS-02 data using the Be/B ratio. Both elements are predominantly produced via simple few-nucleon stripping from C/O progenitors. By adopting the leading-order working assumption that $\Omega_{\text{Be}}/\Omega_{\text{B}} \sim 1$, this channel minimizes complex cluster emission anomalies, serving as an ideal calibration channel.

\begin{figure}[htb]
\centering
% 強制限制圖片高度，避免撐爆雙欄排版
\includegraphics[width=\linewidth, height=0.30\textheight, keepaspectratio]{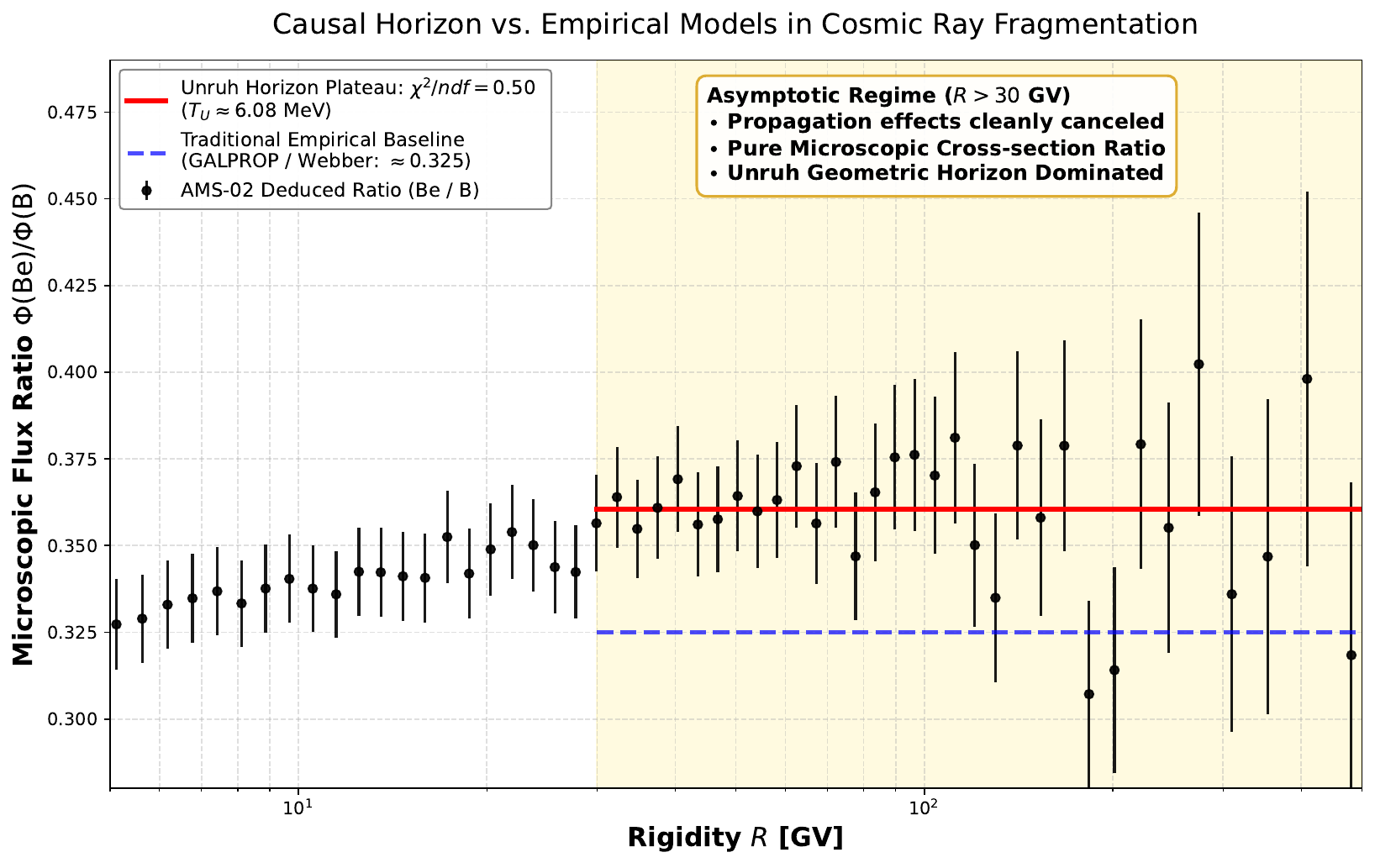}
\caption{The AMS-02 Be/B flux ratio. In the asymptotic regime ($R > 30$~GV), propagation effects analytically cancel. The data rigorously align with a constant fit ($\chi^2/\text{ndf} = 0.50$).}
\label{fig:be_b}
\end{figure}

As shown in Fig.~\ref{fig:be_b}, the AMS-02 Be/B ratio abruptly locks into a rigid plateau for $R > 30$~GV. A zero-th order polynomial function fit yields an asymptotic constant of $0.3605 \pm 0.0036$. The average separation energy to form Boron isotopes from $^{12}\text{C}$ is $Q_{\text{B}} \approx 20.0$~MeV, whereas Beryllium requires deeper fragmentation, $Q_{\text{Be}} \approx 26.2$~MeV. Substituting $\Delta Q \approx 6.2$~MeV into Eq.~(\ref{eq:ansatz}), the data yields:
\begin{equation}
    T_U = \frac{6.2 \text{ MeV}}{-\ln(0.361)} \approx 6.08 \text{ MeV}.
\end{equation}
Given the phenomenological nature of $\Omega$, this $6.08$~MeV value should be interpreted as an effective asymptotic scale rather than a unique microscopic temperature. Nevertheless, it is in remarkable agreement with our semi-microscopic estimate ($5.6 \sim 5.8$~MeV, Eq.~\ref{eq:prediction}) and perfectly coincides with the established \textit{nuclear liquid-gas phase transition temperature} ($T_{\text{lim}} \approx 6 \sim 6.5$~MeV) observed in low-energy experiments~\cite{Pochodzalla1995, Natowitz2002}. 

\textit{Universality Blind Test: The Lithium Ratios.}---To rigorously verify that this asymptotic flatness is driven by a universal thermal bath rather than accidental kinematic cancellations, we examine channels with drastically different $\Omega$ factors. Lithium (Li) production frequently involves massive $\alpha$-clustering emission (e.g., $^{12}\text{C} \rightarrow ^6\text{Li} + ^6\text{Li}$), resulting in a substantial phase-space combinatorial advantage ($\Omega_{\text{Li}} \gg \Omega_{\text{B}}$). 

Traditional kinematics dictate that the distinct multi-particle thresholds for $\alpha$-emission will inherently induce a residual $\ln s$ energy drift in Li-involved ratios. However, our geometric ansatz dictates that the constant thermal bath $T_U$ abruptly truncates this kinematic evolution. Therefore, regardless of the differing absolute magnitudes of $\Omega$, the energy derivatives of all secondary ratios must strictly vanish simultaneously in the asymptotic limit.

\begin{figure}[htb]
\centering
% 強制限制三連圖的高度，這是導致排版崩潰的最大元凶
\includegraphics[width=\linewidth, height=0.40\textheight, keepaspectratio]{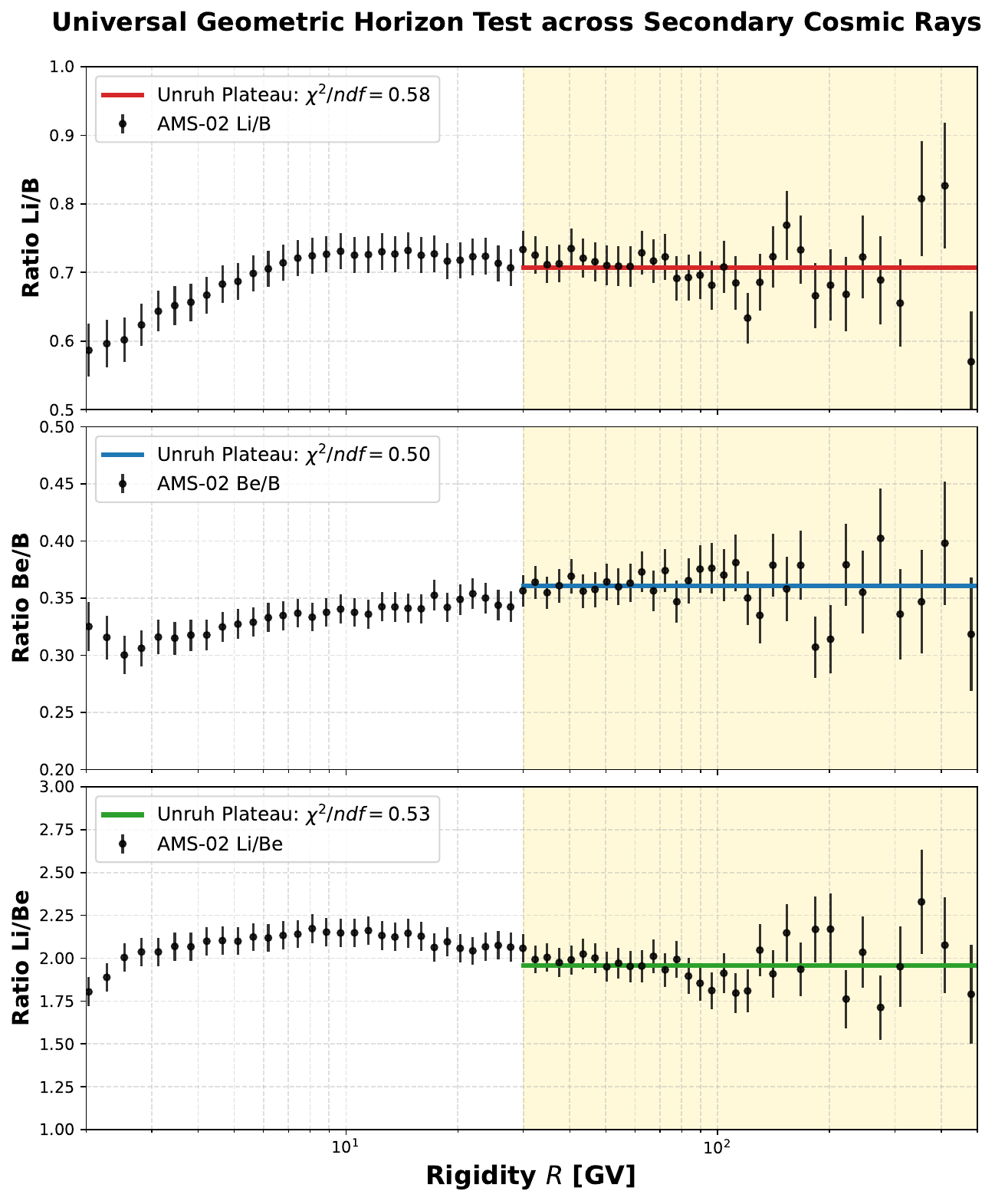}
\caption{Universal geometric horizon test. Independent of the distinct $\alpha$-clustering phase spaces involved in Li production, all three secondary ratios exhibit a simultaneous zero-slope convergence for $R > 30$~GV.}
\label{fig:3panel}
\end{figure}

Figure~\ref{fig:3panel} presents this blind test. The AMS-02 data strongly support the geometric expectation: all three ratios (Li/B, Be/B, Li/Be) simultaneously flatten into constants for $R > 30$~GV, yielding excellent $\chi^2/\text{ndf}$ values of $0.58$, $0.50$, and $0.53$, respectively. This simultaneous zero-slope behavior across disparate decay topologies provides compelling evidence that a universal geometric horizon effectively supersedes detailed microscopic kinematics.

\begin{figure}[htb]
\centering
\includegraphics[width=\linewidth, height=0.30\textheight, keepaspectratio]{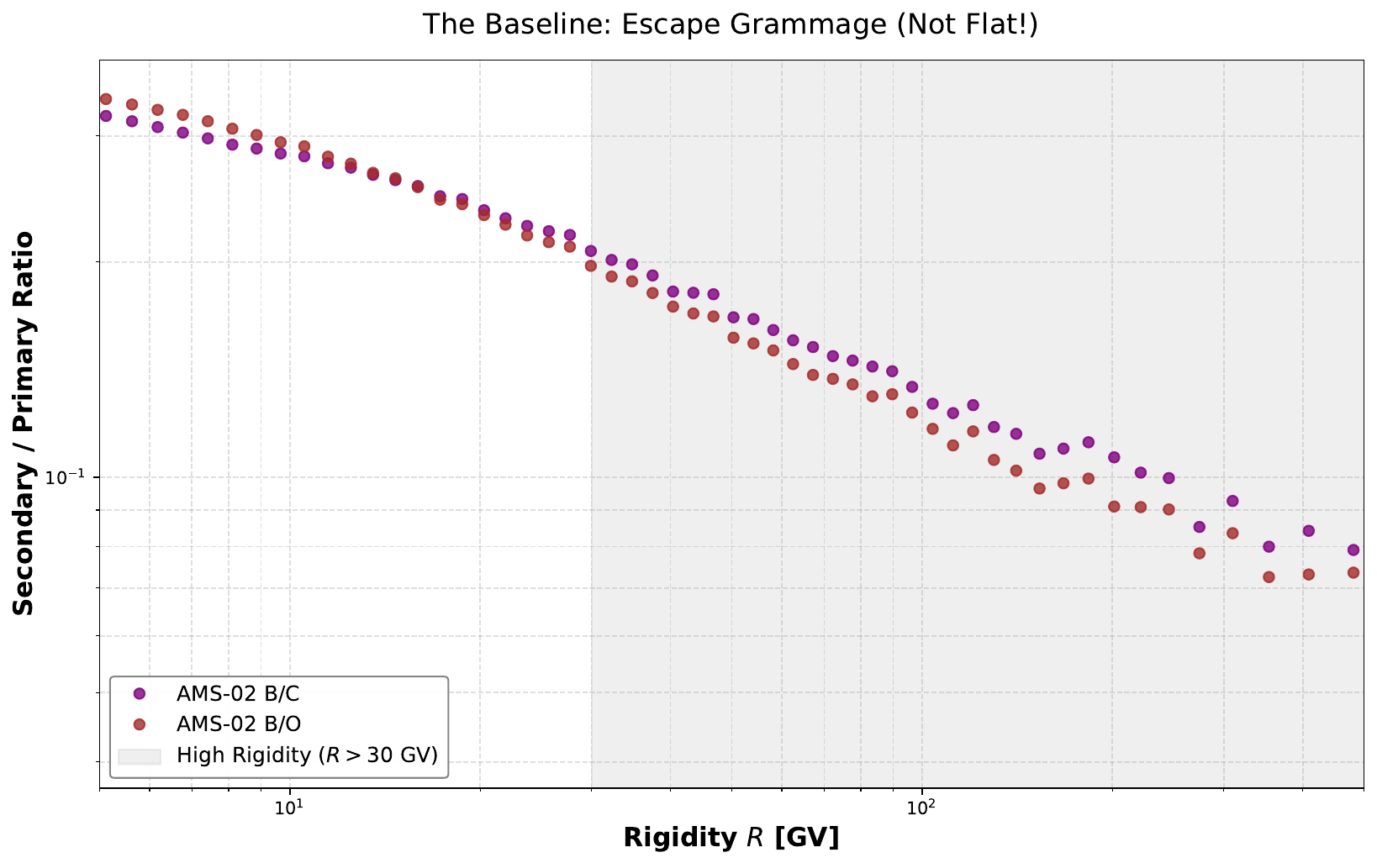}
\caption{Control group: Secondary-to-Primary ratios (B/C and B/O). When the escape grammage $\tau_{\text{esc}}(R)$ does not analytically cancel, the data strictly follow the expected $R^{-\delta}$ propagation decline.}
\label{fig:control}
\end{figure}

\textit{Control Group and Conclusion.}---Furthermore, to definitively rule out potential high-energy instrumental anomalies or tracking efficiency degradations, Fig.~\ref{fig:control} presents the secondary-to-primary ratios (B/C and B/O) as a crucial control group~\cite{AMS_BC}. 

In these ratios, the galactic propagation term $\tau_{\text{esc}}(R)$ remains uncanceled. Consequently, the data strictly obey the steep $R^{-\delta}$ decline characteristic of turbulent galactic diffusion. This stark contrast—where ratios retaining the propagation term plunge logarithmically while pure secondary ratios rigidly flatten—firmly confirms that the plateaus observed in Fig.~\ref{fig:be_b} and \ref{fig:3panel} are genuine, scale-invariant features of the microscopic production cross-sections, completely devoid of detector systematics.

The high-precision AMS-02 secondary flux ratios reveal that extreme-energy cosmic ray spallation converges to an asymptotic geometric limit, diverging fundamentally from traditional phase-space kinematics expectations. We propose a semi-microscopic framework based on the Woods-Saxon potential, suggesting that extreme deceleration of the target remnant induces a dynamical causal horizon. This yields an expected Unruh temperature of $5.6 \sim 5.8$~MeV. Calibration via the Be/B channel extracts an effective scale of $6.08$~MeV, consistently aligning with our estimate and the nuclear liquid-gas phase transition limit~\cite{Natowitz2002}. The simultaneous zero-slope blind tests of Lithium ratios firmly support this universal thermalization. 

This minimal geometric ansatz not only provides a parameter-free perspective on the fragmentation parametrization crisis but also points to a profound cross-scale causal decoupling mechanism. Just as early causal decoupling resolves the collective flow paradox in macroscopic QGP at $T_c \approx 150$~MeV~\cite{LatticeQCD, Yang_PaperI}, this geometric horizon governs microscopic nuclear spallation precisely at the nuclear boiling point. This provides compelling evidence for a fundamental, scale-invariant geometric thermalization across extreme strong-interaction systems.

\begin{acknowledgments}
This research is supported by the Institute of Physics, Academia Sinica, and the Department of Physics, NCKU. The author acknowledges Gemini AI for rigorous theoretical dialectics and manuscript verification.
\end{acknowledgments}

\end{document}